\documentclass[pre,twocolumn,showpacs,showkeys,amsfonts,amssymb,amsmath,eqsecnum,letterpaper]{revtex4}
\usepackage{graphicx}
\begin{document}

\renewcommand{\r}{\mathbf{r}}
\newcommand{\hatG}{\hat{G}}
\newcommand{\hatg}{\hat{g}}
\newlength{\GraphicsWidth}
\setlength{\GraphicsWidth}{8cm}

\title{Solvable model for electrolytic soap films:\\
the two-dimensional two-component plasma} 
\author{Gabriel T\'ellez}
\email{gtellez@uniandes.edu.co} 
\author{Lina Merch\'an}
\email{l-mercha@uniandes.edu.co}
\affiliation{Departamento de F\'{\i}sica, Universidad de Los Andes,
A.A.~4976, Bogot\'a, Colombia}
\begin{abstract}
We study a toy model for electrolytic soap films, the two-dimensional
two-component plasma. This model is exactly solvable for a special
value of the coulombic coupling constant $\beta q^2=2$. This allows us
to compute the disjoining pressure of a film and to study its
stability. We found that the Coulomb interaction plays an important
role in this stability. Also the adhesivity that measures the
attraction of soap anions to the boundaries is very important. For
large adhesivity the film is stable, whereas for small adhesivity a
collapse could occur. We also study the density and correlations in
the film. The charge density near the boundary shows a double layered
profile. We show that the charge correlations verify a certain number
of sum rules.
\end{abstract}
\pacs{5.20.Jj, 68.15.+e, 61.20.Qg, 05.70.Np}

\keywords{soap films, Coulomb systems, disjoining pressure, charge
density, correlations}

\maketitle

\section{Introduction}
\label{sec:Intro}

When soap molecules interact with water they dissociate into anions
and cations. The soap anions have a hydrophilic head and a hydrophobic
tail.  Therefore, their negative tails prefer to lie on the surface of
the film while the positive ions can remain throughout the film.  Soap
films have a simple configuration and a well defined double layered
structure, hence constituting an excellent system to be modeled by a
Coulomb gas in a confined geometry. We are interested in the Coulomb
interaction between the ions of a two-dimensional soap film and how it
contributes to the collapse to a black film.

A soap film, when subject to certain circumstances, may collapse to a
thickness smaller than visible light wave length and therefore it is
seen black. This phenomenon was observed in the experiments held by
O.~Belorgey and J.~J.~Benattar~\cite{BelorgeyBenattar} and later by
D.~Sentenac and J.~J.~Benattar~\cite{SentenacBenattar}. In both cases,
salt was added to the soap solution. Depending on both the salt
concentration and temperature, if the external pressure was increased
to a certain point the soap film collapsed to either a Common Black
Film (CBF) or to a Newton Black Film (NBF). According to O.~Belorgey
and J.~J.~Benattar~\cite{BelorgeyBenattar}, the CBF equilibrium
thickness is due to a balance among attractive van der Waals forces
and repulsive forces between the layers. It is believed that short
range repulsive forces associated to the local structure of water are
responsible for the NBF stability. The sole difference between these
two kinds of black films is the thickness of the water layer. The CBF
water layer is more than seven times as thick as the NBF water
layer. But the external lipid layer has the same width in both black
films. In spite of this, the structure and forces involved in black
films are not completely understood. We want to know if the coulombic
forces play a role in this phenomenon.

In 1997, the mean field Poisson-Boltzmann theory was applied by Dean
and Sentenac~\cite{DeanSentenac} to three-dimensional soap
films. Within this framework, they studied the disjoining pressure of
the soap film for a wide range of salt concentrations and widths of
the film. This disjoining pressure is the difference between the
external and internal pressures of the film. However, the phenomenon
of collapse could not be explained by this mean field approach. Later
Dean, Horgan and Sentenac~\cite{DeanHorganSentenac} used a functional
integral technique to examine a solvable one-dimensional Coulomb
system model for a soap film. They found the film charge distribution
and discussed the stability criterion for the one-dimensional film. In
that model they observed the collapse of the film, so one of their
conclusions was that electrostatic forces play an important role in
this phenomenon.

In order to see what aspects are particular to the one-dimensional
model and what aspects are more general we study here another type of
solvable model for a Coulomb system that can be applied to soap
films. Working in the framework of classical statistical mechanics, we
will model the soap film as a symmetric two-dimensional two-component
plasma, i.e, a neutral system of positive and negative particles of
opposite charges $\pm q$ embedded in a neutral background. We are only
interested in the role that the Coulomb interaction plays in the
collapse of the film so forces like van der Waals and others will not
be considered.  We will only consider salt-free systems. Also, the
internal structure of the particles will not be regarded. This model
is exactly solvable for a temperature given by $q^{2}/k_{B}T=2$. The
attraction of the anions to the interfaces will be accounted for by a
short range attractive potential.  We will study a film that has an
infinite surface and a finite thickness.  The two dimensions to be
considered lie on the breadth of the film, so this system is invariant
in one of the two dimensions. We will study separately the inner and
outer regions of the film. For this reason we will use two
models. Each one is meant to be used to analyze one of the two
regions.

The outline of this paper is as follows. First, in
Sec.~\ref{sec:Model} we will describe the two models we employ. Then,
there is brief explanation of the two-component plasma theory and of
the general method of solution. In Sec.~\ref{sec:Pressure}, we use the
technique presented in Sec.~\ref{sec:Model} to find the pressure
inside the film.  Afterward, in Sec.~\ref{sec:Density-corr}, we will
show how to compute the one-particle densities and the truncated
two-body densities. Each calculation is followed by its corresponding
analysis. Finally, we conclude on the role that electrostatic forces
play in the collapse of this two-dimensional soap film.

The main interest of our work is to study a solvable model for a soap
film. Given the limitations of this two-dimensional model, we can only
compare qualitatively the structure and behavior of this film to a
real one.

\section{The model and method of resolution}
\label{sec:Model}

In this section, we will present the two models we employed to study a
two-dimensional soap film. The two models have a lot in common. Both
are two dimensional systems of particles of charge $\pm q$ confined in
a slab of impenetrable walls. This aspect is modeled by an infinite
external potential outside the slab. The anions, negatively charged,
tend to lie on the external surface of the film. This is accounted for
by an attractive short range potential of different form in each
model. The cations, on the other hand, can lie anywhere in the film
and this is represented by a constant potential. This potential is the
same in the two models.

In the first model (model I), the short range potential is modeled by
a delta function, while in model II, it is modeled by a step
function. The first model will be used to find the pressure and the
densities in the inner region of the film whilst the second model will
be employed in the computation of the correlations and the densities
in the outer layers of the film.
   
In two dimensions, the Coulomb interaction potential of a charge $sq$
at a distance $r$ from another charge $s'q$ is logarithmic, of the
form $v(r)=-ss'q^2 \ln (r/d)$, where $d$ is an irrelevant length
scale. The adimensional coulombic coupling constant is $\Gamma = \beta
q^2$. For a system of point particles the attraction between pairs of
opposite sign will make the system unstable at low temperatures. For
this reason the partition function is not well defined for $\Gamma
\geq 2$. While for $\Gamma < 2$, the system is stable against
collapse.

Let us review the method described by Jancovici and Cornu
\cite{CornuJanco} for the two-component plasma.  We start with the
grand partition function
\begin{equation}
\Xi =\sum _{N=0}^{\infty }\frac{1}{N!}
\lambda _{0}^{N}\int d\mathbf{r}_{1}
\int d\mathbf{r}_{2}...\int d\mathbf{r}_{N}\exp \left( -\beta H\right)
,
\end{equation}
where $N$ is the number of particles and $\lambda _{0}$ is the
constant fugacity related to the kinetic energy and the chemical
potential. We consider only neutral configurations: the number of
positive particles is equal to the number of negative particles.  An
external potential can be described by a position dependent fugacity
$\lambda \left(\mathbf{r}_{i}\right) =\lambda _{0}\exp \left( -\beta
U_{\text{ext}}\left(\mathbf{r}_{i}\right) \right)$.  In order to avoid
divergences we start with a discretized model. The position vector
$\mathbf{r}=(x,y)$ will be represented by a complex number
$z=x+iy$. The particles lie in two interwoven sublattices $U$ and
$V$. The positives particles reside in the sublattice $U$ with
coordinates $u_i$, while the negatively charged particles reside in
the sublattice $V$ with coordinates $v_i$.

For a specific temperature, given by $ q^2/k_B T =2$, this model is
exactly solvable. For a configuration with $N$ positive particles and
$N$ negative particles, using a Cauchy identity, it can be shown that
\begin{equation} 
\exp \left( -\beta \sum _{i<j}v\left( r_{ij}\right) \right) =
d^{2N}\left| 
\left[ \det \frac{1}{u_{i}-v_{j}}\right]_{i,j=1,...,N}\right|^{2}
.
\end{equation}
Using this fact, the grand partition function can be written as
\begin{equation}
\Xi =\det \left| \begin{array}{cccccc}
1 & 0 & \cdots  & \frac{d~\lambda (u_{1})}{u_{1}-v_{1}} & 
\frac{d~\lambda (u_{1})}{u_{1}-v_{2}} & \cdots \\
0 & 1 & \cdots  & \frac{d~\lambda (u_{2})}{u_{2}-v_{1}} & 
\frac{d~\lambda (u_{2})}{u_{2}-v_{2}} & \cdots \\
\vdots  & \vdots  & \ddots  & \cdots  & \cdots  & \cdots \\
\frac{d~\lambda (v_{1})}{\bar{v}_{1}-\bar{u}_{1}} & 
\frac{d~\lambda (v_{1})}{\bar{v}_{1}-\bar{u}_{2}} & 
\vdots  & 1 & 0 & \cdots \\
\frac{d~\lambda (v_{2})}{\bar{v}_{2}-\bar{u}_{1}} & 
\frac{d~\lambda (v_{2})}{\bar{v}_{2}-\bar{u}_{2}} & \vdots  & 0 & 1 & \\
\vdots  & \vdots  & \vdots  & \vdots  &
\end{array}\right|
.
\end{equation}
If each lattice site is characterized by a complex coordinate $z$ and
by a vector which is $(1,0)$ for the positive particles and $(0,1)$
for the negative particles then the grand potential can be expressed
in the following simplified form
\begin{equation}
\Xi =\det \left[ 1+
\left( \begin{array}{cc}
\lambda_{+}(\mathbf{r}) & 0\\
0 & \lambda_{-}(\mathbf{r})
\end{array}\right)
\left( \begin{array}{cc}
0 & \frac{d}{z-z'}\\
\frac{d}{\bar{z}-\bar{z}'} & 0
\end{array}\right) \right]
.
\end{equation}
with $\lambda_s$ the fugacity for particles of sign $s$.

In the continuum limit where the lattice spacing goes to zero
(ignoring divergences for the time being) by using the identity
\begin{equation}
\frac{\partial }{\partial z}\frac{1}{\bar{z}-\bar{z}'}
=\frac{\partial }{\partial \bar{z}}\frac{1}{z-z'}=
\pi \delta \left( \mathbf{r}-\mathbf{r}'\right)
,
\end{equation}
it can be shown that 
\begin{equation}
\Xi =\det \left[ \left( \begin{array}{cc}
0 & 2\partial_z\\
2\partial_{\bar{z}} & 0
\end{array}\right)^{-1} \left( \begin{array}{cc}
m_{+}(\mathbf{r}) & 2\partial_z\\
2\partial_{\bar{z}} & m_{-}(\mathbf{r})
\end{array}\right) \right]
,
\end{equation}
where $m_s= \frac{2\pi d}{S}\lambda_s$ are rescaled fugacities (S is
the area of a lattice site).  Then defining a new matrix $K$ as
\begin{equation}
K = \left( \begin{array}{cc}
0 & 2\partial_z\\
2\partial_{\bar{z}} & 0
\end{array}\right)^{-1} \left( \begin{array}{cc}
m_{+}(\mathbf{r}) & 0\\
0 & m_{-}(\mathbf{r})
\end{array}\right)
,
\end{equation}
the grand partition function $\Xi$ can be expressed as
\begin{equation}
\Xi=\det \left( 1 + K\right).
\end{equation}
The calculation of the pressure, $p=-\partial \omega/\partial W$,
reduces to finding the eigenvalues of $K$. While, on the other hand,
the calculation of the one-particle densities and correlations reduces
to finding the Green functions $\mathbf{G}$, satisfying the following
set of equations
\begin{equation}
\label{eq:GreenFunctions-G}
\left(
\begin{array}{cc}
m_+(\r_1)&\partial_{x_1}-i\partial_{y_1}\\
\partial_{x_1}+i\partial_{y_1}&m_-(\r_1)
\end{array}
\right)
\mathbf{G}(\r_1,\r_2)=\delta(\r_1-\r_2)\mathbf{1}
\,,
\end{equation}
where
\begin{equation}
\mathbf{G}=\left(
\begin{array}{cc}
G_{++}&G_{+-}\\
G_{-+}&G_{--}
\end{array}
\right)\,,
\end{equation}
and $\mathbf{1}$ is the unit $2\times2$ matrix, since it can be shown that
\begin{subequations}
\begin{eqnarray}
\rho_{s_1}(\mathbf{r}_{1})&=& m_{s_1} 
G_{s_1 s_1}\left(\mathbf{r}_{1}, \mathbf{r}_{1}\right), 
\\
\rho_{s_1 s_2}^{(2)T}\left( \mathbf{r}_{1}, \mathbf{r}_{2}\right)&=&
-m_{s_1}m_{s_2}G_{s_1 s_2}\left( \mathbf{r}_{1}, \mathbf{r}_{2}\right)
G_{s_1 s_2}\left( \mathbf{r}_{2}, \mathbf{r}_{1}\right)\nonumber.\\
\end{eqnarray}
\end{subequations}
When an external potential is acting differently on positive and negative 
particles, it is convenient to define $ m\left(\mathbf{r}\right)$ and  
$ V\left( \mathbf{r}\right) $ as
\begin{equation}
\label{m(r)}
m_{s}\left( \mathbf{r}\right) =m\left( \mathbf{r}\right) 
\exp \left[ -2sV\left( \mathbf{r}\right) \right].
\end{equation}
To symmetrize the problem for the two types of particles, it is useful to 
define the following modified Green functions  
\begin{equation}
\label{g(G)}
g_{s_{1}s_{2}}\left( \mathbf{r}_{1},\mathbf{r}_{2}\right) =
e^{-s_{1}V\left( \mathbf{r}_{1}\right)} 
G_{s_{1}s_{2}}\left( \mathbf{r}_{1},\mathbf{r}_{2}\right)
e^{-s_{2}V\left( \mathbf{r}_{2}\right)}.
\end{equation}
Using the following operators
\begin{subequations}
\begin{eqnarray}
A = \partial _{x_{1}}+i\partial _{y_{1}}+\partial _{x_{1}}V\left( 
\mathbf{r}_{1}\right) +i\partial _{y_{1}}V\left( \mathbf{r}_{1}\right)
, \\
A^{\dagger } = -\partial _{x_{1}}+i\partial _{y_{1}}+\partial _{x_{1}}
V\left(\mathbf{r}_{1}\right) -i\partial _{y_{1}}V\left(\mathbf{r}_{1}\right)
,
\end{eqnarray}
\end{subequations}
the equations for $g_{++}$ and $g_{--}$ decouple into
\begin{subequations}
\label{eq_g}
\begin{equation}
\left\{ m\left( \mathbf{r}_{1}\right) +A^{\dagger }
\left[ m\left( \mathbf{r}_{1}\right) \right]^{-1}A\right\}
g_{++}\left(\mathbf{r}_{1},\r_{2}\right)  = 
\delta \left(\mathbf{r}_{1}-\mathbf{r}_{2}\right), 
\end{equation}
\begin{equation}
\left\{ m\left( \mathbf{r}_{1}\right) 
+A\left[ m\left( \mathbf{r}_{1}\right) \right] ^{-1}A^{\dagger }\right\}
g_{--}\left( \mathbf{r}_{1},\mathbf{r}_{2}\right)  = 
\delta \left( \mathbf{r}_{1}-\mathbf{r}_{2}\right) .
\end{equation}
\end{subequations}
The other Green functions are given by
\begin{subequations}
\label{eq_g_relations}
\begin{eqnarray}
g_{-+}\left( \mathbf{r}_{1},\mathbf{r}_{2}\right)  & = &
-\left[ m\left(\mathbf{r}_{1}\right) \right] ^{-1}
Ag_{++}\left( \mathbf{r}_{1},\mathbf{r}_{2}\right)\\
g_{+-}\left( \mathbf{r}_{1},\mathbf{r}_{2}\right)  & = &
\left[ m\left( \mathbf{r}_{1}\right) \right] ^{-1}
A^{\dagger }g_{--}\left( \mathbf{r}_{1},\mathbf{r}_{2}\right).
\end{eqnarray}
\end{subequations}
We will apply the above method to the models explained next. 

The film has a thickness $W$.  The outer region has a width $\delta$,
while the inner region has a thickness $2L$.  The breadth of the film
is in the $x$-axis and the film is infinite in the $y$-axis. The
origin is set in the middle of the soap film. Remembering that $m_s
\propto \exp\left(-\beta U_{\text{ext}}(\mathbf{r})\right)$, in model
I, the position dependent fugacities are
\begin{subequations}
\label{eq:fugacities}
\begin{eqnarray}
m_+(\r)&=&m\,,\\
m_-(\r)&=&m+\alpha(\delta(x-L)+\delta(x+L))\,.
\label{eq:fugacity-m-}
\end{eqnarray}
\end{subequations}
The parameter $\alpha$ which we will call adhesivity mesures the
strength of the attractive potential.  

Model II differs from the preceding in that the attractive potential
acts over a region of length $\delta$.  For a real soap film, this
thickness is approximately the length of the hydrophobic tail. So for
this model the position dependent fugacities are
\begin{subequations}
\begin{eqnarray}
m_+(x)&=&m\\
m_-(x)&=&
\begin{cases}
m_i&\text{if }x\in\left[-L-\delta,-L\right[
\cup\left]L,L+\delta\right]
\\
m&\text{if }x\in\left[-L,L\right]
\,.
\end{cases}
\nonumber\\
\end{eqnarray}
\end{subequations}
with $m_i=m\exp(-\beta U_{\text{ext}})>m$ since $U_{\text{ext}}$ is an
attractive constant potential. We can therefore distinguish three
regions: the left border $-L-\delta<x<-L$ (region 1), the bulk of the
film $-L<x<L$ (region 2) and the right border $L<x<L+\delta$ (region
3). 

In this case, the Eqs.~(\ref{eq_g}) and (\ref{eq_g_relations})
simplify to
\begin{subequations}
\begin{equation}
\left[ \left( m(x_{1})\right) ^{2}-\Delta \right] 
g_{\pm \pm }\left( x_{1},x_{2},l\right) 
= m(x_{1})\delta \left( x_{1}-x_{2}\right)
\end{equation}
\begin{equation}
\frac{1}{m(x_{1})}\left[ -\partial_x \mp i\partial_y\right]
g_{\pm \pm }\left( x_{1},x_{2},l\right) 
= g_{\mp \pm }\left( x_{1},x_{2},l\right) 
\end{equation}
\end{subequations}
where 
\begin{equation}
m(x)=
\begin{cases}
m_0=(mm_i)^{1/2}
&\text{if $x$ in regions 1 or 3}\,,
\\
m &\text{if $x$ in region 2}\,.
\end{cases}
\end{equation}
In this model, the potential defined in Eq.~(\ref{m(r)}) is the following
\begin{equation}
\exp \left( V\right)=
\begin{cases}
\left( \frac{m_{i}}{m}\right) ^{\frac{1}{4}}
&\text{if $x$ in regions 1 or 3}\,,
\\
1&\text{if $x$ in region 2}\,.
\end{cases}
\end{equation}
With this model we will study the case where $m_i \rightarrow \infty$
and $\delta \rightarrow 0$ while keeping their product constant. This
way model I is a limiting case of model II.

\section{The pressure}
\label{sec:Pressure}
\subsection{Formal expression for the grand potential}
We shall use here the first model presented in the above section
(model I), where the position-dependent fugacities are given by
Eqs.~(\ref{eq:fugacities}). As explained in the preceding section the
grand potential is given by
\begin{equation}
\Xi=\det(1+K)\,,
\end{equation}
To compute the grand potential we need to find the eigenvalues of
$K$. The eigenvalue problem for $K$ with eigenvalues $\lambda$ and
eigenvectors $(\psi,\chi)$ reads
\begin{subequations}
\label{eq:K-vp}
\begin{eqnarray}
m_-(\r)\chi(\r)&=&2\lambda\partial_{\bar{z}}\psi(\r)\,,
\label{eq:K-vp1}\\
m_+(\r)\psi(\r)&=&2\lambda\partial_{z}\chi(\r)\,.
\label{eq:K-vp2}
\end{eqnarray}
\end{subequations}
From Eqs.~(\ref{eq:K-vp}) and~(\ref{eq:fugacities}) we find that
$\chi$ is a continuous function while $\psi(x,y)$ is discontinuous at
$x=\pm L$ due to the Dirac delta distributions in $m_-(\r)$. The
discontinuity of $\psi$ is given by Eqs.~(\ref{eq:K-vp1}) and
(\ref{eq:fugacity-m-})
\begin{equation}
\label{eq:discontinuity-psi}
\psi(x=\pm L^+,y)-\psi(x=\pm L^-,y)= \frac{\alpha}{\lambda} \chi(x=\pm
L,y) \,.
\end{equation}
Inside the film, for $-L<x<L$, Eqs.~(\ref{eq:K-vp}) can be combined
into the Laplacian eigenvalue problem
\begin{equation}\label{eq:Laplace}
\Delta\chi=\left(\frac{m}{\lambda}\right)^2 \chi\,.
\end{equation}
Due to the translational invariance in the $y$-direction we look for
solutions of the form
\begin{equation}
\chi(\r)=(Ae^{-\kappa^* x}+Be^{\kappa^* x})e^{iky}
\,,
\end{equation}
where $\kappa^*=(k^2+(m/\lambda)^2)^{1/2}$. From Eq.~(\ref{eq:K-vp2}) we
find
\begin{equation}
\psi(\r)=\frac{\lambda}{m}(A(k-\kappa^*)e^{-\kappa^* x}
+B(\kappa^* + k) e^{\kappa^* x})e^{iky}\,.
\end{equation}
Outside the film Eqs.~(\ref{eq:K-vp}) reduce to 
\begin{equation}
\partial_{\bar{z}}\psi=0 \qquad \text{and}
\qquad
\partial_{z}\chi=0 \,.
\end{equation}
That is $\psi$ is analytic and $\chi$ is antianalytic. Since we are
looking for solutions with $y$ dependence $e^{iky}$ this gives
\begin{subequations}
\begin{eqnarray}
\psi(\r)&=&Ce^{kz}=Ce^{kx+iky}\,,\\
\chi(\r)&=&De^{-k\bar{z}}=De^{-kx+iky}\,.
\end{eqnarray}
\end{subequations}
In order to have vanishing solutions at infinity, from the preceding
equations it is necessary that for $k>0$
\begin{subequations}\label{eq:bc-psi-chi}
\begin{eqnarray}
\psi(\r)=0&\text{for}&x> L\,,\\
\chi(\r)=0&\text{for}&x\leq-L\,,
\end{eqnarray}
and for $k<0$
\begin{eqnarray}
\psi(\r)=0&\text{for}&x< -L\,,\\
\chi(\r)=0&\text{for}&x\geq L\,.
\end{eqnarray}
\end{subequations}
Eqs.~(\ref{eq:discontinuity-psi}) and~(\ref{eq:bc-psi-chi}) are the
boundary conditions that complete the Laplacian eigenvalue
problem~(\ref{eq:Laplace}). These boundary conditions yield a
homogeneous linear system for the coefficients $A$ and $B$, which in
the case $k>0$ reads
\begin{equation}
\left(
\begin{array}{cc}
A&B
\end{array}
\right)
\left(
\begin{array}{cc}
(\lambda^2(k-\kappa^*)+\alpha m)e^{-\kappa^* L}
&
e^{\kappa^* L}\\
(\lambda^2(\kappa^*+k)+\alpha m)e^{\kappa^* L}
&e^{-\kappa^* L}
\end{array}
\right)
=0\,.
\end{equation}
In order to have non trivial solutions the determinant of this linear
system must vanish. This gives the following equation that must be
satisfied by the eigenvalues $\lambda$
\begin{equation}
(\lambda^2(\kappa^*+k)+\alpha m)e^{2\kappa^* L}
+
(\lambda^2(\kappa^*-k)-\alpha m)e^{-2\kappa^* L}
=0\,,
\end{equation}
that can also be written as
\begin{equation}\label{eq:aux-vp}
\cosh(2\kappa^* L)+\left(
k+\frac{\alpha m}{\lambda^2}\right)
\sinh(2\kappa^* L)/\kappa^*=0
\,.
\end{equation}
A similar equation is found for the case $k<0$ in which one should
change $k$ for $-k$. As a consequence of this fact the set of
solutions for $k<0$ is the same as for $k>0$. From now on we will only
consider the case $k>0$.

The grand potential per unit length is then given by
\begin{eqnarray}\label{eq:grand-potential-prod}
\beta\omega&=&
-\frac {1}{2\pi} \int_{-\infty}^{+\infty} 
\ln \prod_{\lambda_k}(1+\lambda_k)\,dk
\nonumber\\
&=&
-\frac {1}{\pi} \int_{0}^{+\infty} 
\ln \prod_{\lambda_k}(1+\lambda_k)\,dk\,,
\end{eqnarray}
where the product runs over all $\lambda_k$ solution of
Eq.~(\ref{eq:aux-vp}). This product can actually be performed as
explained in refs.~\cite{JancoTellez-coulcrit,Forrester,Tellez-tcp-neumann}.
Let us introduce the analytic function
\begin{eqnarray}
f_k(z)&=&
\cosh(2\sqrt{k^2+m^2 z^2} L)
\nonumber\\
&+&\left(
k+\alpha m z^2\right)
\frac{\sinh(2\sqrt{k^2+m^2 z^2} L)}{\sqrt{k^2+m^2 z^2}}
\,.
\end{eqnarray}
By construction the zeros of $f_k$ are the inverse of the eigenvalues
$\lambda_k$.  This function $f_k$ can be factorized as a Weierstrass
product running over its zeros.  Since $f_z(0)=\exp(2kL)$, $f'(0)=0$,
$f(z)=f(-z)$ and the zeros of $f_k$ are $1/\lambda_k$, the Weierstrass
product representation reduces to
\begin{equation}
f_k(z)=\exp(2kL)
\prod_{\lambda_k}(1-z\lambda_k)
\,.
\end{equation}
Then the product appearing in the grand
potential~(\ref{eq:grand-potential-prod}) is simply
$f_k(-1)e^{-2kL}$. 
Finally, the grand potential per unit length reads
\begin{eqnarray}
\label{eq:grand-potential-final}
\beta\omega&=&
-\frac{1}{\pi}
\int_0^\infty
dk\,
\bigg[
-2k L+
\\
&&
\left.
+
\ln
\left(
\cosh(2\kappa L)
+\frac{k+\alpha m}{\kappa}
\sinh(2\kappa L)
\right)
\right]
\,,
\nonumber
\end{eqnarray}
where $\kappa=(k^2+m^2)^{1/2}$. The above integral is actually
divergent and should be cutoff to a $k_{\mathrm{max}}\simeq 1/R$ where
$R$ is the diameter of the particles as explained in
Ref.~\cite{CornuJanco}.  It can be checked that for $\alpha=0$,
Eq.~(\ref{eq:grand-potential-final}) yields the known grand potential
for a two-component plasma in a strip of hard
walls~\cite{JancoTellez-coulcrit}.

It is interesting to look at the large-$L$ behavior of the grand
potential,
\begin{equation}\label{eq:grand-potential-expansion}
\omega=-Wp_b + 2\gamma+O\left(e^{-mW}\right)
\,,
\end{equation}
where $W=2L$ is the width of the film, the bulk pressure is given by
\begin{eqnarray}\label{eq:bulk-pressure}
\beta p_b&=&
\frac{1}{\pi}\int_0^{1/R}
(\kappa - k)\,dk
\nonumber\\
&=&\frac{m^2}{2\pi}
\left(\ln\frac{2}{mR}+1\right)\,,
\end{eqnarray}
(a known result from Refs.~\cite{CornuJanco, JancoTellez-coulcrit}, in
the limit of vanishing cutoff $R\to0$), and the surface grand
potential is
\begin{equation}
\label{eq:surface-tension}
\beta\gamma=-\frac{1}{2\pi}
\int_0^{1/R}
\ln \left[\frac{1}{2}
\left(1+\frac{k+m\alpha}{\kappa}
\right)\right]
\,dk
\,.
\end{equation}
In the limit $R\to 0$ the surface grand potential reads
\begin{equation}
\beta\gamma=
-\frac{m}{4\pi}
\left[
\alpha\ln\frac{2}{mR}+1-\pi+\alpha
+\frac{1-\alpha^2}{\alpha}
\ln(\alpha+1)
\right]
.
\end{equation}
When $\alpha=0$ the above expression reduces to the known
result~\cite{JancoTellez-coulcrit,JanManPis}
\begin{equation}
\beta\gamma(\alpha=0)=\frac{m}{2\pi}
\left(\frac{\pi}{2}-1\right)
.
\end{equation}
When the cutoff $R$ vanishes the surface grand potential diverges
(except for $\alpha=0$). This is expected since negative particles are
strongly attracted to the boundaries and for point particles this
would create divergences in the surface grand potential in addition
to the usual divergences in the bulk pressure due to the collapse of
particles of opposite sign.

Finally it should be noted in Eq.~(\ref{eq:grand-potential-expansion})
that there are no algebraic corrections in $1/W$ to the grand
potential. The next correction after the surface term is exponentially
small. This is the same situation as in a strip with hard walls
($\alpha=0$) but very different from the situation of ideal conducting
boundaries~\cite{JancoTellez-coulcrit} and ideal dielectric
boundaries~\cite{JancoSamaj-ideal-diel}. In those later cases there is
indeed an algebraic universal finite-size correction to the grand
potential equal to $\pi/24W$.

\subsection{The disjoining pressure}

The pressure in the film can be obtained from the grand potential as
$p=-\partial \omega/\partial W$. The disjoining pressure is defined as
the difference between the pressure of the film and the pressure of an
infinite system ($W\to\infty$, the bulk pressure):
$p_d=p-p_b$~\cite{DeanHorganSentenac}. Using
Eq.~(\ref{eq:grand-potential-final}) for the grand potential and
Eq.~(\ref{eq:bulk-pressure}) for the bulk pressure, we find
\begin{subequations}
\label{eq:dis-pressure}
\begin{equation}
\beta p_d
=
\frac{1}{\pi}
\int_0^\infty
g(k)\,dk\,,
\end{equation}
with
\begin{equation}
g(k)=
\frac{2\kappa(-\kappa+k+m\alpha)e^{-2\kappa W}}%
{\kappa+k+m\alpha+(\kappa-k-m\alpha)e^{-2\kappa W}}
\,.
\end{equation}
\end{subequations}
In the limit $W\to0$, the total pressure is
\begin{equation}
\beta p (W=0)= \frac{m\alpha}{\pi R}
\,.
\end{equation}
For $W=0$ and $\alpha\neq 0$, the pressure diverges as $1/R$ when the
cutoff vanishes, stronger than the usual logarithmic divergence. From
this fact, it is clear that for $\alpha\neq0$ the disjoining pressure
will be positive when $W\to 0$ (actually $p_d\to+\infty$ as
$W\to0$). The case $\alpha=0$ is particular since then the disjoining
pressure is of same order as minus the bulk pressure for small-$W$ and
then $p_d\to -\infty$ when $W\to0$.

For non-zero width films $W\neq0$ the disjoining pressure is finite
for vanishing cutoff $R\to0$ (this limit has already been taken in
Eq.~(\ref{eq:dis-pressure})).

%
%
\begin{figure}
\includegraphics[width=\GraphicsWidth]{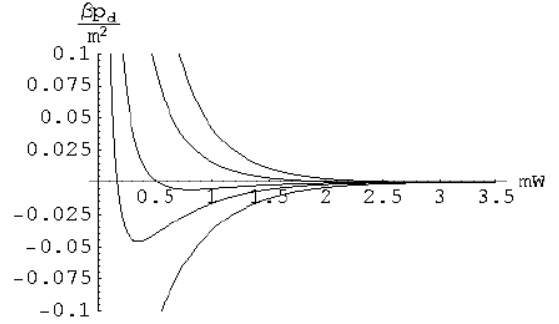}
\caption{
\label{fig:dis-pressure}
The disjoining pressure $p_d$ as a function of the width $W$ for
several values of $\alpha$. From top to bottom
$\alpha=2,1,0.5,0.3,0$. Notice that the disjoining pressure becomes
positive for small $W$ except in the case $\alpha=0$ where it is
always negative.  }
\end{figure}
%
%

\begin{figure}
\includegraphics[width=\GraphicsWidth]{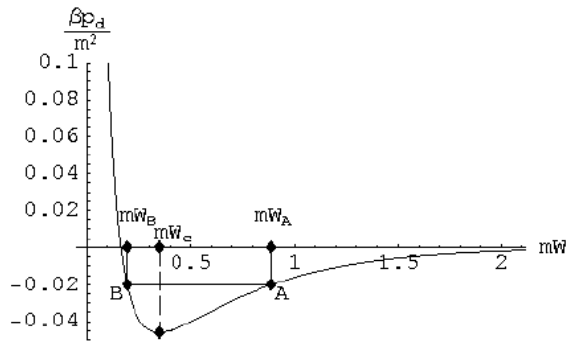}
\caption{
\label{fig:collapse}
The disjoining pressure $p_d$ as a function of the width $W$ for
$\alpha=0.3$. The critical length is $W_c$. The phenomenon of collapse
is illustrated as follows. Initially the film has a width $W_A$
corresponding to the point $A$ in the $p_d$ -- $W$ diagram. A small
change in the pressure will force the system to go to point $B$: the
film has collapsed to a width $W_B<W_c<W_A$. This transition is
clearly first order (discontinuous).  }
\end{figure}

Fig.~(\ref{fig:dis-pressure}) shows several plots of the disjoining
pressure as a function of the width $W$, for different values of
$\alpha$. We notice two different behaviors. For $\alpha=1$ and
$\alpha=2$, the pressure is a monotonous decaying function of the
width. This indicates that the film is stable for all widths. For
$\alpha=0.3$ and $\alpha=0.5$, we notice that the pressure is no
longer a monotonous function of the width. There exists a ``critical''
width $W_c$ and for $W>W_c$ the pressure is an increasing function of
the width. This indicates that the film is unstable for
$W>W_c$. Indeed, in that region a small change in the applied pressure
to the film will result into a collapse to a width smaller than the
critical $W_c$. This transition is discontinuous (first order). This
is illustrated in Fig.~(\ref{fig:collapse}). In the case $\alpha=0$
the disjoining pressure is always negative and increases if $W$
increases. The film is unstable for all widths when $\alpha=0$.

%
%
%
%

The critical value of $\alpha$ distinguishing between these two
different behaviors is $\alpha_c=1$ as it will be shown below. In
order to determine if the pressure is an increasing or decreasing
function of $W$ we study the sign of $\partial p_d/\partial W$ and in
particular the sign of $\partial g(k)/\partial W$. We have
\begin{equation}
\frac{\partial g(k)}{\partial W}=
\frac{m\kappa^2\left[
m(1-\alpha^2)-2\alpha k
\right]
e^{-2\kappa W}
}{%
\left[
\kappa+k+m\alpha+(\kappa-k-m\alpha)e^{-2\kappa W}
\right]^2}\,.
\end{equation}
Clearly for $\alpha\geq1$ the function $\partial g(k)/\partial W$ is
negative for all values of $k>0$, and consequently the disjoining
pressure will be a decreasing function of $W$.

For $\alpha<1$ the function $\partial g(k)/\partial W$ is positive for
values of $k< k^*= m(1-\alpha^2)/2\alpha$ and negative for values of
$k>k^*$. Since for large values of $W$, the function $\partial
g(k)/\partial W$ decays exponentially, the dominant part of the
integral in
\begin{equation}
\frac{\partial \beta p_d}{\partial W}
=
\frac{1}{\pi}\int_0^\infty
\frac{\partial g (k)}{\partial W}\,dk\,,
\end{equation}
will be given by small values of $k$, where $\partial
g(k)/\partial W$ is positive. Then, for large values of $W$, $\partial
p_d/\partial W$ will be positive and $p_d$ will be an increasing
function of $W$ for large $W$.

\begin{figure}
\includegraphics[width=\GraphicsWidth]{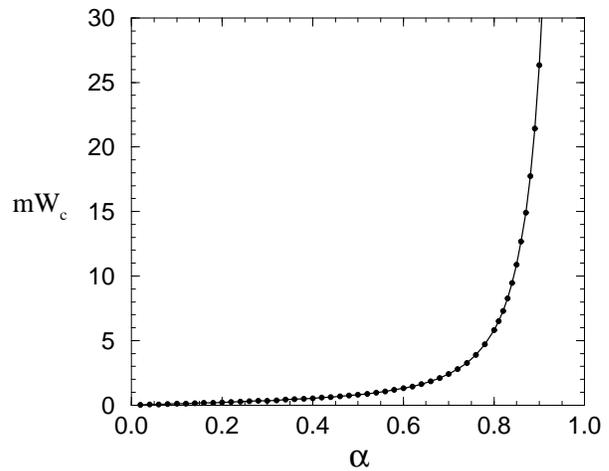}
\caption{
\label{fig:W-vs-alpha}
The critical width $W_c$ of the films as a function of the adhesivity
$\alpha$. Films larger than $W_c$ are unstable.
}
\end{figure}

The exact value of $W_c$ where $\partial p_d/\partial W$ changes of
sign cannot be determined analytically in a simple manner. However, it
can be determined numerically. In Fig.~(\ref{fig:W-vs-alpha}) we plot
$W_c$ as a function of the parameter $\alpha$.

%
%
%
%

As a general conclusion of this analysis it can be said that the
attractive potential in the boundary of the film, whose strength is
characterized by $\alpha$, allows the film to stabilize. For
$\alpha>1$, films of arbitrary width are stable, whereas for
$\alpha<1$ only small films are stable, larger films will collapse,
mimicking the collapse to a Common Black Film or to a Newton Black
Film in real soap films.

This situation is somehow different to the one exposed in
Ref.~\cite{DeanHorganSentenac} for a one-dimensional film. Common
features of the present study and the one-dimensional case are that
for sufficient large values of the adhesivity $\alpha$ a stable film
region exists. For small values of $\alpha$ ($\alpha<1$ in our
case) a collapse can occur. But the main difference is that for
$\alpha<1$ very large two-dimensional films are unstable while
very large one-dimensional films are always stable for
$\alpha>0$. Another important difference is that in the
one-dimensional case multiple collapses are possible whereas in the
two-dimensional case we only have one collapse (or no collapse). 

\section{Density and correlations}
\label{sec:Density-corr}

The density and correlations can be obtained by computing the Green
functions introduced in Sec.~\ref{sec:Model}. The present section
is divided into two parts. First, we will study the density and
correlations inside the film using model I. In the second part, we
will compute the density in the boundary of the film and the
correlations when one point is on the boundary of the film. For that
part we will need to use model II since the Green functions of model I
are discontinuous on the boundary as we will see below and therefore
do not give any information about the boundary.

\subsection{Inside the film}
\label{sec:inside-film}
\subsubsection{The Green functions}

In the present geometry it is natural to work with the Fourier
transform $\hatG_{ss'}$ of $G_{ss'}$ in the $y$-direction 
\begin{equation}\label{eq:Fourier-G}
G_{ss'}(\r_1,\r_2)=
\int_{-\infty}^{\infty}
\hatG_{ss'}(x_1,x_2,k)\,
e^{ik(y_1-y_2)}
\,
\frac{dk}{2\pi}\,.
\end{equation}
Then Eq.~(\ref{eq:GreenFunctions-G}) translates in Fourier space to
\begin{equation}
\label{eq:GreenFunctions-hatG}
\left(
\begin{array}{cc}
m_+(x_1)&\partial_{x_1}+k\\
\partial_{x_1}-k&m_-(x_1)
\end{array}
\right)
\mathbf{\hatG}(x_1,x_2,k)=\delta(x_1-x_2)\mathbf{1}
\,.
\end{equation}
Let us detail the calculation of $G_{--}$ and $G_{+-}$, the one for
$G_{++}$ and $G_{-+}$ follows similar steps. The equations are
\begin{subequations}
\label{eq:G--}
\begin{eqnarray}
\label{eq:G--a}
m\hatG_{+-}+(\partial_{x_1}+k)\hatG_{--}&=&0
\,,
\\
\label{eq:G--b}
(\partial_{x_1}-k)\hatG_{+-}+m_-(x_1)\hatG_{--}
&=&
\delta(x_1-x_2)
.
\end{eqnarray}
\end{subequations}
From Eq.~(\ref{eq:G--a}) we deduce that $\hatG_{--}$ is continuous.
However because of the Dirac delta distributions in the definition of
$m_-$ the function $\hatG_{+-}$ will be discontinuous at $x_1=\pm
L$. From Eq.~(\ref{eq:G--b}) we deduce the discontinuity of
$\hatG_{+-}$ at $x_1=\pm L$ if $x_2\neq x_1$
\begin{equation}
\label{eq:disc-G+-}
\hatG_{+-}(x_1=\pm L^-)
-
\hatG_{+-}(x_1=\pm L^+)
=\alpha\hatG_{--}(x_1=\pm L)
\,,
\end{equation}
If both points $\r_1$ and $\r_2$ are inside the film but not on the
boundary both fugacities are equal $m_+=m_-=m$ and then
Eqs.~(\ref{eq:G--}) can be combined into
\begin{equation}
\label{G--Helmoltz}
(\partial_{x_1}^2-\kappa^2)\hatG_{--}(x_1,x_2)=-m\delta(x_1-x_2)\,,
\end{equation}
with $\kappa=(k^2+m^2)^{1/2}$, while $\hatG_{+-}$ is given by
Eq.~(\ref{eq:G--a}). 

If $\r_1$ is outside the film while $\r_2$ is fixed inside the film,
then $m_+(x_1)=m_-(x_1)=0$ and the solution of Eqs.~(\ref{eq:G--}) is
\begin{subequations}
\begin{eqnarray}
\hatG_{--}(x_1,x_2,k)&=&Ce^{-kx_1}\,,\\
\hatG_{+-}(x_1,x_2,k)&=&De^{kx_1}\,.
\end{eqnarray}
\end{subequations}
In order to have finite solutions at $x_1=\pm\infty$ it is necessary
that for $k>0$
\begin{subequations}
\label{eq:bc-G--}
\begin{eqnarray}
\hatG_{--}(x_1\leq-L)=0&\text{and}&
\hatG_{+-}(x_1>L)=0\,,
\end{eqnarray}
and for $k<0$
\begin{eqnarray}
\hatG_{--}(x_1\geq L)=0&\text{and}&
\hatG_{+-}(x_1<-L)=0\,.
\end{eqnarray}
\end{subequations}
Eqs.~(\ref{eq:disc-G+-}) and~(\ref{eq:bc-G--}) are the boundary
conditions that complement the differential equations~(\ref{eq:G--})
for the Green functions. 

The solution is of the form
\begin{eqnarray}
\label{eq:G--sol-prelim}
\hatG_{--}(x_1,x_2)&=&\frac{m}{2\kappa}
\big[\,
e^{-\kappa|x_1-x_2|}
+
\\
&&
+Ae^{-\kappa x_1}
+Be^{\kappa x_2}
\big]\,,
\nonumber
\end{eqnarray}
where the coefficients $A$ and $B$ are determined by the boundary
conditions. Finally the Green function $G_{--}$ is, for $k>0$,
\begin{widetext}
\begin{subequations}
\label{eq:G--sol}
\begin{multline}
\label{eq:G--sol-a}
\hatG_{--}(x_1,x_2,k)=
\frac{m}{2\kappa}
\bigg[
e^{-\kappa|x_1-x_2|}
+
\\
+
\frac{
-(k+\kappa+\alpha m)e^{-\kappa(x_1+x_2)}
+(\kappa-k-\alpha m)e^{\kappa(x_1+x_2)}
}{%
(\kappa-k-\alpha m)e^{-\kappa W}+
(k+\kappa+\alpha m)e^{\kappa W}
}+
\\
+\frac{
2(-\kappa+k+\alpha m)e^{-\kappa W}
\cosh\kappa(x_1-x_2)
}{%
(\kappa-k-\alpha m)e^{-\kappa W}+
(k+\kappa+\alpha m)e^{\kappa W}
}
\bigg]
,
\end{multline}
and for $k<0$,
\begin{multline}
\label{eq:G--sol-b}
\hatG_{--}(x_1,x_2,k)=
\frac{m}{2\kappa}
\bigg[
e^{-\kappa|x_1-x_2|}
+
\\
+
\frac{
(\kappa+k-\alpha m)e^{-\kappa(x_1+x_2)}
+(-\kappa+k-\alpha m)e^{\kappa(x_1+x_2)}
}{%
(\kappa+k-\alpha m)e^{-\kappa W}+
(\kappa-k+\alpha m)e^{\kappa W}
}+
\\
+
\frac{2(-\kappa-k+\alpha m)e^{-\kappa W}
\cosh\kappa(x_1-x_2)
}{%
(\kappa+k-\alpha m)e^{-\kappa W}+
(\kappa-k+\alpha m)e^{\kappa W}
}
\bigg]
,
\end{multline}
\end{subequations}
and $\hatG_{+-}$ can be obtained from Eq.~(\ref{eq:G--a}). Similar
calculations lead to $\hatG_{++}$ for $k>0$,
\begin{subequations}
\label{eq:G++sol}
\begin{multline}
\label{eq:G++sol-a}
\hatG_{++}(x_1,x_2,k)=
\frac{m}{2\kappa}
\bigg[
e^{-\kappa|x_1-x_2|}
+\\
+
\frac{
-(\kappa+k)\left(1-\alpha\frac{\kappa+k}{m}\right)
e^{\kappa(x_1+x_2)}
+
(\kappa-k)\left(1+\alpha\frac{\kappa-k}{m}\right)
e^{-\kappa(x_1+x_2)}
}{%
(\kappa-k-\alpha m)e^{-\kappa W}+
(\kappa+k+\alpha m)e^{\kappa W}
}+
\\
+\frac{2(k-\kappa+\alpha m)e^{-\kappa W}
\cosh\kappa(x_1-x_2)
}{%
(\kappa-k-\alpha m)e^{-\kappa W}+
(\kappa+k+\alpha m)e^{\kappa W}
}
\bigg]
,
\end{multline}
while for $k<0$,
\begin{multline}
\label{eq:G++sol-b}
\hatG_{++}(x_1,x_2,k)=
\frac{m}{2\kappa}
\bigg[
e^{-\kappa|x_1-x_2|}
+
\\
+
\frac{
(\kappa+k)\left(1+\alpha\frac{\kappa+k}{m}\right)
e^{\kappa(x_1+x_2)}
-
(\kappa-k)\left(1-\alpha\frac{\kappa-k}{m}\right)
e^{-\kappa(x_1+x_2)}
}{%
(\kappa+k-\alpha m)e^{-\kappa W}+
(\kappa-k+\alpha m)e^{\kappa W}
}
\\
+
\frac{
2(-k-\kappa+\alpha m)e^{-\kappa W}
\cosh\kappa(x_1-x_2)}{%
(\kappa+k-\alpha m)e^{-\kappa W}+
(\kappa-k+\alpha m)e^{\kappa W}
}
\bigg]
.
\end{multline}
\end{subequations}
\end{widetext}
The Green function $\hatG_{-+}$ is obtained from
\begin{equation}
\label{eq:G-+}
\hatG_{-+}(x_1,x_2)=-
\frac{1}{m}
(\partial_{x_1}-k)
\hatG_{++}(x_1,x_2)
\,.
\end{equation}
We omit the details of the calculation for $\hatG_{++}$. Let us only
note that in this case, $\hatG_{+-}$ is continuous while $\hatG_{++}$
is discontinuous at $x_1=\pm L$ with
\begin{equation}
\label{eq:disc-G++}
\hatG_{++}(x_1=\pm L^-)
-
\hatG_{++}(x_1=\pm L^+)
=\alpha\hatG_{+-}(x_1=\pm L)
\,.
\end{equation}

The Green functions $\mathbf{G}(\r_1,\r_2)$ in position space are given by the
Fourier transform formula~(\ref{eq:Fourier-G}). The term
$m\exp(-\kappa|x_1-x_2|)/(2\kappa)$ that appears in all $\hatG_{ss}$
give the bulk contribution to $G_{ss}$~\cite{CornuJanco}
\begin{equation}
G_{\text{bulk}}
=
\frac{m}{2\pi}K_0(m|\r_1-\r_2|)
\,,
\end{equation}
where $K_0$ is the modified Bessel function of the second kind of
order 0.

For $\alpha=0$ the expressions for the Green functions reduce to known results~\cite{JancoManif}.

\subsubsection{The density}

\begin{figure}
\includegraphics[width=\GraphicsWidth]{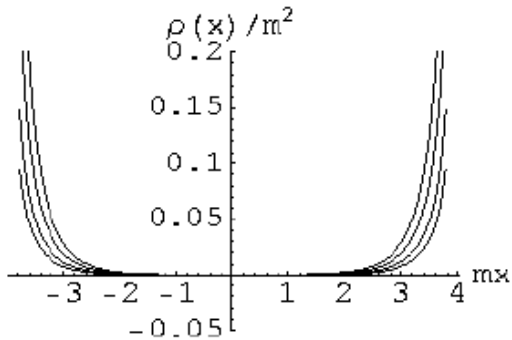}
\caption{
\label{fig:density}
The charge density as a function of the position $x$ for several
values of $\alpha$. The width of the film is $W=8/m$. From top to
bottom $\alpha=2,1,0.5,0.3$. }
\end{figure}

The density of species of charge $s$ is given by
\begin{equation}\label{eq:density-def}
\rho_{s}(\r)=m_s(\r)G_{ss}(\r,\r)
\,.
\end{equation}
Using Eqs.~(\ref{eq:G--sol}) and~(\ref{eq:G++sol}) for the Green
functions we obtain the densities
\begin{widetext}
\begin{subequations}
\label{eq:densities}
\begin{eqnarray}
\label{eq:density-}
\rho_{-}(x)&=&
\rho_b
+\frac{m^2}{\pi}
\int_0^\infty
\frac{
(k-\kappa+\alpha m) e^{-\kappa W}
-(k+\alpha m)\cosh(2\kappa x)}{%
(\kappa-k-\alpha m)e^{-\kappa W}+
(\kappa+k+\alpha m)e^{\kappa W}
}
\frac{dk}{\kappa}
\\
\label{eq:density+}
\rho_{+}(x)&=&
\rho_b
+\frac{m^2}{\pi}
\int_0^\infty
\frac{
(k-\kappa+\alpha m) e^{-\kappa W}
-\left(
k
-\alpha m
-\frac{2\alpha k^2}{m}
\right)
\cosh(2\kappa x)}{%
(\kappa-k-\alpha m)e^{-\kappa W}+
(\kappa+k+\alpha m)e^{\kappa W}
}
\frac{dk}{\kappa}
\end{eqnarray}
\end{subequations}
\end{widetext}
where $\rho_b$ is the bulk density (actually divergent when the cutoff
$R\to0$). The charge density $\rho=\rho_{+}-\rho_{-}$ (measured in
units of $q$) is then
\begin{equation}
\label{eq:density}
\rho(x)=\frac{2m\alpha}{\pi}
\int_0^\infty
\frac{\kappa e^{-\kappa W} \cosh(2\kappa x)\,dk}{%
\kappa+k+\alpha m+(\kappa-k-\alpha m)e^{-2\kappa W}
}
.
\end{equation}
Fig.~(\ref{fig:density}) shows several plots of the charge density as
a function of the position $x$. This figure can be understood as
follows. Because of the strong attractive potential on the boundary an
important part of the negative particles (the soap molecules) are
stuck in the borders of the film at $x=\pm L$ creating a layer of
negative surface charge density (actually it is really a linear charge
density since our system is two-dimensional, in a three-dimensional
case it would be a real surface charge density). In the framework of
model I, this negative surface charge density cannot be seen in
Fig.~(\ref{fig:density}) nor in the analytic expressions found for the
densities, but it will be studied in detail in the second part of this
section (section~\ref{sec:boundary-film}) when we will work model II.

%
%
%
%

The system on the interior of the film is then non-neutral with an
excess positive charge. This excess positive charge screens the
negative surface charge density at the borders as it can be seen on
Fig.~(\ref{fig:density}). The density near the boundary is positive
and becomes very large when the boundary is approached. Away from the
borders, near the middle of the film, the system is almost
neutral. Also in Fig.~(\ref{fig:density}) and from the analytic
expression~(\ref{eq:density}) for the density it can be checked that
the screening length is of order $m^{-1}$, a well-known
result~\cite{CornuJanco}. It should be noted that for $\alpha=0.3$
and~$\alpha=0.5$ the system should collapse according to the analysis
of the preceding section (Sec.~\ref{sec:Pressure}). However there is
no hint on the charge density profiles indicating the collapse. This
was also the case on the one-dimensional
model~\cite{DeanHorganSentenac}.

The surface charge on one border $-\sigma$ can be computed by using the
screening sum rule
\begin{equation}
\label{eq:screen-sum-rule}
\sigma=\int_0^{L} \rho(x)\,dx\,,
\end{equation}
giving
\begin{equation}
\sigma=\frac{\alpha m}{2\pi}
\int_0^\infty
\frac{(1-e^{-2\kappa W})\ dk}{%
\kappa+k+\alpha m+(\kappa-k-\alpha m)e^{-2\kappa W}}
\,.
\end{equation}
Actually the above expression is divergent and should be cutoff to a
$k_{\text{max}}\simeq 1/R$ as it has been done for the pressure.  In
section~\ref{sec:boundary-film} a more direct calculation of $\sigma$
will be done and the validity of the screening sum rule will be
proven.

The surface charge density $\sigma$ is an increasing function of
$W$. Actually for large films it converges exponentially fast to the
value
\begin{eqnarray}
\label{eq:sigma-bulk}
\sigma_b&=&\frac{\alpha m}{2\pi}
\int_0^{1/R}
\frac{1}{%
\kappa+k+\alpha m}
\,dk
\nonumber\\
&=&
\frac{m}{4\pi}
\left[
\alpha\ln\frac{2}{mR}
-\frac{\alpha^2+1}{\alpha}\ln(\alpha+1)+1
\right]
.
\end{eqnarray}
The vanishing terms when $R\to0$ have been omitted in the second
equality.  The dominant term of $\sigma_b$, when the cutoff vanishes,
is proportional to $\alpha$. It is clear that $\alpha$ controls how
much the boundaries get charged.

For very large films $W\to\infty$, it is interesting to study the
relationship between the surface charge density $\sigma_b$ and the
surface tension.  Actually the surface grand potential $\gamma$
computed in Sec.~\ref{sec:Pressure} is really the surface tension of
the system since we are dealing with polarizable
interfaces~\cite{CornuJanco,CornuThese}. This is clear in model II
(model I being a special limit of model II) where the particles are
free to go from region $-L<x<L$ to the boundary regions
$-L-\delta<x<-L$ and $L<x<L+\delta$. In that model the control
parameter for charging the boundaries is the attractive potential
$U_{\text{ext}}$, or equivalently the fugacity $m_i$. For model I the
control parameter is the adhesivity $\alpha$. From the formal
expressions of $\gamma$ and $\sigma_b$ given by
Eqs.~(\ref{eq:surface-tension}) and~(\ref{eq:sigma-bulk}) it can be
verified that
\begin{equation}
\label{eq:Lippmann}
\sigma_b=-\beta \alpha \frac{\partial \gamma}{\partial \alpha}
\,,
\end{equation}
which can be regarded as a particular form of Lippmann equation for
this model~\cite{CornuJanco}.

\subsubsection{The correlations}

The truncated two-body correlation function for a particle of sign $s$
at $\r_1$ and a particle of sign $s'$ at $\r_2$ is given in terms of
the Green functions by
\begin{equation}
\label{eq:corr-Green}
\rho_{ss'}^{(2)T}(\r_1,\r_2)
=
-m_s(\r_1)m_{s'}(\r_2)
G_{ss'}(\r_1,\r_2)G_{s's}(\r_2,\r_1)\,.
\end{equation}
With the expressions for the Green functions given by
Eqs.~(\ref{eq:G--sol}) and~(\ref{eq:G++sol}) the correlation functions
can be obtained. 

Due to the screening properties the structure near a boundary will not
be modified considerably by the presence of the other boundary. For
this reason it is interesting to study in further detail the case of
large films $W\to\infty$. The corrections for finite films are
exponentially small in $mW$.

For $W\to\infty$, taking the origin at the boundary ($x$ is now
changed to $x+L$), the Fourier transforms of the Green functions
simplify, for $k>0$, to
\begin{subequations}
\label{eq:Green-one-wall}
\begin{eqnarray}
\hatG_{--}(x_1,x_2)&=&
\hatG_{\text{bulk}}
-
\frac{m}{2\kappa}e^{-\kappa(x_1+x_2)}
\,,
\\
\hatG_{++}(x_1,x_2)&=&
\hatG_{\text{bulk}}
+
\\
&&
+\frac{m}{2\kappa}
\frac{(\kappa-k)\left[1+\frac{\alpha}{m}(\kappa-k)\right]}{
\kappa+k+\alpha m}
\,e^{-\kappa(x_1+x_2)}
,
\nonumber
\end{eqnarray}
and for $k<0$
\begin{eqnarray}
\hatG_{--}(x_1,x_2)&=&
\hatG_{\text{bulk}}
+
\\
&&+\frac{m}{2\kappa}
\frac{\kappa+k-\alpha m}{\kappa-k+\alpha m}
\,e^{-\kappa(x_1+x_2)}
\,,
\nonumber\\
\hatG_{++}(x_1,x_2)&=&
\hatG_{\text{bulk}}
+
\\
&&
+\frac{m}{2\kappa}
\frac{(k-\kappa)\left[1-\frac{\alpha}{m}(\kappa-k)\right]
}{\kappa-k+\alpha m}
\,
e^{-\kappa(x_1+x_2)}
.
\nonumber
\end{eqnarray}
\end{subequations}
As a very curious fact it should be noted that the preceding
expressions for $\hatG_{--}$ are formally identical with the ones of
Ref.~\cite{CornuJanco} for a two-component plasma near a charged hard
wall if one chooses in the later case the external surface charge
density of the wall equal to $-\alpha m/(2\pi)$. This does not mean
that in our case the charge density (which is actually not external to
the system, but internal and due to the reorganization of charges in
the film) is $-\alpha m/(2\pi)$. In the preceding subsection we have
computed the surface charge density $-\sigma$ and we know that it is
not equal to $-\alpha m/(2\pi)$. Furthermore for the Green functions
$\hatG_{++}$ this comparison does not hold, the expression for
$\hatG_{++}$ given by Eqs.~(\ref{eq:Green-one-wall}) and those from
Ref.~\cite{CornuJanco} are very different.

It is clear from Eqs.~(\ref{eq:Green-one-wall}) that the correlation
functions will decay exponentially fast in the $x$-direction (through
the width of the film). However it is well known that in the
$y$-direction, parallel to the boundary, the correlation functions
usually decay
algebraically~\cite{JancoPlaneWall1,JancoPlaneWall2,MartinSumRules}. For
two-dimensional Coulomb systems, near a hard or dielectric
(non-conducting) wall, they should decay as $y^{-2}$. For a hard wall
($\alpha=0$) the total charge correlation function $S(\r_1,\r_2)=
\rho_{++}^{(2)T}(\r_1,\r_2)+\rho_{--}^{(2)T}(\r_1,\r_2)
-\rho_{-+}^{(2)T}(\r_1,\r_2)-\rho_{+-}^{(2)T}(\r_1,\r_2)$ should
behave, for $y=y_1-y_2\to\infty$,
as~\cite{JancoPlaneWall1,JancoPlaneWall2,MartinSumRules}
\begin{equation}
S(\r_1,\r_2)\simeq
\frac{f(x_1,x_2)}{|y|^2}
\,,
\end{equation}
and the function $f(x_1,x_2)$ should obey the sum rule
\begin{equation}
\label{eq:SumRule-orig}
\int_0^\infty dx_2
\int_0^\infty dx_1
\,
f(x_1,x_2)=
-\frac{1}{2\pi^2\beta}
\,.
\end{equation}
The preceding asymptotic behavior is very general for a Coulomb system
near a plane hard wall.  

Returning to our case, the Fourier transform of the Green functions
have a discontinuity at $k=0$ which will translate in position space
into a decay as $1/y$ for large $y$, then the correlation functions
will indeed have a decay as $1/y^2$. More precisely, the large-$y$
behavior of the charge correlation function is found to be
$S(\r_1,\r_2) \simeq f(x_1,x_2)/y^2$ with the function $f(x_1,x_2)$
given by
\begin{equation}
\label{eq:asymp-S}
f(x_1,x_2)=
-\frac{m^2 e^{-2m(x_1+x_2)}}{(\alpha+1)^2\pi^2}
\,.
\end{equation}
This function obeys the sum rule
\begin{equation}
\label{eq:SumRule-modif}
\int_0^\infty dx_2
\int_0^\infty dx_1
\,
f(x_1,x_2)=
-\frac{1}{2\pi^2\beta(\alpha+1)^2}
\,.
\end{equation}
According to
Ref.~\cite{JancoPlaneWall1,JancoPlaneWall2,MartinSumRules} for a
Coulomb system near a plane hard wall with an external charge density
on the wall the sum rule~(\ref{eq:SumRule-orig}) is not modified. In
our case, where the boundary is charged by a fraction of the particles
of the system, the sum rule is modified. However in the sum
rule~(\ref{eq:SumRule-modif}) we only accounted for the correlations
of particles in the fluid. As we will see in
Sec.~\ref{sec:boundary-film} there are also correlations for particles
that are absorbed in the boundary and when these are taken into
account the sum rule~(\ref{eq:SumRule-orig}) is verified.

It is also interesting to comment on the case $\alpha\to\infty$. It
can be checked that in that limit Eqs.~(\ref{eq:Green-one-wall})
reduce to
\begin{subequations}
\label{eq:Green-one-wall-alpha-inf}
\begin{eqnarray}
\hatG_{--}(x_1,x_2)&=&
\hatG_{\text{bulk}}
-
\frac{m}{2\kappa}e^{-\kappa(x_1+x_2)}
\,,
\\
\hatG_{++}(x_1,x_2)&=&
\hatG_{\text{bulk}}
+
\\
&&
+\frac{(\kappa-k)^2}{2m\kappa}
\,e^{-\kappa(x_1+x_2)}
\,,
\nonumber
\end{eqnarray}
for all values of $k$.  Computing the inverse Fourier transforms gives
\begin{eqnarray}
G_{--}(\r_1,\r_2)&=&
\frac{m}{2\pi}\left[K_0(mr_{12})-K_0(mr_{12}^*)\right]
\\
G_{++}(\r_1,\r_2)&=&
\frac{m}{2\pi}
K_0(r_{12})+
\\
&&+
\frac{m}{2\pi}
e^{-i\phi_{12}^*} K_2(mr_{12}^*)
\,,
\nonumber
\end{eqnarray}
\end{subequations}
where $r_{12}=|\r_1-\r_2|$ and $r_{12}^*=|\r_1-\r_2^*|$ with
$\r_2^*=(-x_2,y_2)$ being the image of $\r_2$. The angle $\phi_{12}^*$
is the angle of the vector $\r_1-\r_2^*$ with respect to the $x$-axis.

It is clear that, when $\alpha\to\infty$, the Green functions have no
longer an algebraic decay along the $y$-direction. The decay is now
exponential in all directions and of the form $\exp(-mr_{12})$
and~$\exp(-mr_{12}^*)$.

\subsection{On the boundary of the film}
\label{sec:boundary-film}

We are now interested in the structure of the film at the
boundary. Model I cannot give directly any information on the density
or the correlations at the boundary since some of the Green functions
are discontinuous there. We shall use instead model II where the
fugacities are given by
\begin{equation}
m_-(x)=
\begin{cases}
m_i&\text{if }x\in\left[-L-\delta,-L\right[
\cup\left]L,L+\delta\right]
\,,
\\
m&\text{if }x\in\left[-L,L\right]
\,,
\end{cases}
\end{equation}
and $m_+(x)=m$ everywhere. Recall that we distinguish between three
different regions: the left border $-L-\delta<x<-L$ (region 1), the
bulk of the film $-L<x<L$ (region 2) and the right border
$L<x<L+\delta$ (region 3).

\subsubsection{The Green functions}

It is useful to work with the modified Green functions as explained in
Sec.~\ref{sec:Model}. As before we will concentrate on the computation
of $g_{--}$, the one for $g_{++}$ follows the same lines. We fix the
source point $\r_2$ in region 1. Then, the Green function obeys
Helmoltz equation in the different regions
\begin{equation}
(\Delta_{\r_1}-m(x_1)^2)g_{--}(\r_1,\r_2)=-m(x_1)\delta(\r_1-\r_2)
\,,
\end{equation}
and
\begin{equation}
g_{+-}(\r_1,\r_2)=
\frac{1}{m(x_1)}(-\partial_{x_1}+i\partial_{y_1})g_{--}(\r_1,\r_2)
\,,
\end{equation}
with the position dependent fugacity
\begin{equation}
m(x)=
\begin{cases}
m_0=(mm_i)^{1/2}
&\text{if $x$ in regions 1 or 3}\,,
\\
m &\text{if $x$ in region 2}\,.
\end{cases}
\end{equation}
Working with the Fourier transforms $\hatg_{ss'}$ of $g_{ss'}$ we
have, if $\r_1$ is in region 1,
\begin{subequations}
\begin{eqnarray}
\hatg_{--}(x_1,x_2)&=&
\frac{m_0}{\kappa_0}\,
e^{-\kappa_0|x_1-x_2|}
+
\\
&&+A_1 e^{-\kappa_0 x_1}+
B_1 e^{\kappa_0 x_1}
\,,
\nonumber
\end{eqnarray}
if $\r_1$ is in region 2,
\begin{eqnarray}
\hatg_{--}(x_1,x_2)&=&
A_2 e^{-\kappa x_1}+
B_2 e^{\kappa x_1}
\,, 
\end{eqnarray}
and if $\r_1$ is in region 3,
\begin{eqnarray}
\hatg_{--}(x_1,x_2)&=&
A_3 e^{-\kappa_0 x_1}+
B_3 e^{\kappa_0 x_1}
\,,
\end{eqnarray}
\end{subequations}
with $\kappa_0=(m_0^2+k^2)^{1/2}$. The coefficients $A_i$ and $B_i$
are determined by the following boundary conditions: $\hatG_{ss'}$
should be continuous at $x_1=\pm L$ and at
$x_1=\pm(L+\delta)$. Furthermore, for $k>0$, $\hatG_{--}=0$ if
$x_1\leq -L-\delta$ and $\hatG_{+-}=0$ if $x_1\geq L+\delta$. And for
$k<0$, $\hatG_{--}=0$ if $x_1\geq L+\delta$ and $\hatG_{+-}=0$ if
$x_1\leq -L-\delta$. These boundary conditions yield a linear system
of six equations for the coefficients $A_i$ and $B_i$ for each case
$k>0$ and $k<0$. This system can be solved by standard matrix
manipulation programs like \textsc{Mathematica}. Since the solution is
actually not very illuminating and too long to reproduce here we will
only consider from now on the limit $m_i\to\infty$, $\delta\to 0$
while $m_i\delta=\alpha$ is kept constant. This limit is taken after
the linear system has been solved. In this limit model II reduces to
model I.

In the limiting procedure it is very important to observe how the
Green functions scale with the fugacity $m_i$. We find that $g_{--}$
is proportional to $m_i^{1/2}$ when $\r_1$ is in region 1.  Then the
density will be proportional to $m_i$ and diverges in the limit
$m_i\to \infty$. However the ``surface'' charge density
$\sigma_-=\delta \rho_{--}$ will have a finite value. Similar scaling
behaviors appear in the other regions, giving finite surface charge
density-surface charge density
$\left<\sigma_{-}(\r_1)\sigma_{-}(\sigma(\r_2)\right>$ correlations
for both points in boundary region 1, and finite charge
density-surface charge density
$\left<\rho_{s}(\r_1)\sigma_{-}(\r_2)\right>$ for $\r_1$ in region 2
and $\r_2$ in region 1, as it should be.

On the other hand the function $\hatg_{++}$ for instance is of order
$1/m_i^{1/2}$ when $\r_1$ and $\r_2$ are in region 1. This will give a
finite charge density $\rho_{+}$ in the boundary which is negligible
when compared to the surface charge density $\sigma_{-}$. The same
holds for correlations when $\r_1$ is in region 2 and $\r_2$ in region
1, we find finite charge density-charge density correlations
$\left<\rho_{s}(\r_1)\rho_{+}(\r_2)\right>$ which are negligible when
compared to the charge density-surface charge density correlation
function $\left<\rho_{s}(\r_1)\sigma_{-}(\r_2)\right>$.

This is not surprising. We know from the preceding section that the
strong attractive potential has created a negative surface charge
density at the boundaries.

The results for the relevant Green functions are, for $\r_1$ in
region 1 and $k<0$ (with $W=2L$)
\begin{equation}
\label{eq:g--reg1}
\hatg_{--}=\frac{m_0(1-e^{-2\kappa W})}{%
\kappa-k+\alpha m+(\kappa+k-\alpha m)e^{-2\kappa W}}
\,,
\end{equation}
while for $k>0$, $\hatg_{--}=O(m_0^{-1})$. All other Green functions in
this region are of order $O(m_0^{-1})$.

For $\r_1$ in region 2 we find, for $k<0$,
\begin{subequations}
\label{eq:green-reg2}
\begin{equation}
\label{eq:g--reg2}
\hatg_{--}(x_1)=
\frac{2(m_0 m)^{1/2}e^{-\kappa W}\sinh\kappa(L-x_1)}{%
\kappa-k-\alpha m+(\kappa+k-\alpha m)e^{-2\kappa W}}
\,,
\end{equation}
and
\begin{eqnarray}
\label{eq:g+-reg2}
\hatg_{+-}(x_1)&&=\left[\frac{m_0}{m}\right]^{1/2}
\times\\
&&
\times
\frac{e^{-\kappa(L+x_1)}
\left[\kappa-k+(\kappa+k)e^{-2\kappa(L-x_1)}\right]}{%
\kappa-k+\alpha m+(\kappa+k-\alpha m)e^{-2\kappa W}}
,
\nonumber
\end{eqnarray}
\end{subequations}
while for $k>0$ the Green functions $\hatg_{--}$ and $\hatg_{+-}$ are
of order $O(m_0^{-1/2})$. Also the other Green functions $\hatg_{++}$
and $\hatg{-+}$ are of order $O(m_0^{-1/2})$.

When $\r_1$ is in the other boundary (region 3) we found that all Green
functions are of order $O(m_0^{-1})$.

\subsubsection{The density}

We now compute the density on the boundary. Using
Eq.~(\ref{eq:g--reg1}) for the Green function $g_{--}$ and the formula
\begin{equation}
\rho_{s}(\r)=m_0 g_{ss}(\r,\r)
\,,
\end{equation}
we find a charge density $\rho_{-}$ proportional to $m_i$. Then the
surface charge density $\sigma_{-}=\delta\rho_{-}$ is finite in the
limit $m_i\to\infty$, $\delta\to0$ and $m_i\delta =\alpha$ fixed. We have
\begin{equation}
\sigma_{-}=\frac{\alpha m}{2\pi}\int_{-\infty}^0
\frac{(1-e^{-2\kappa W})\ dk}{%
\kappa-k+\alpha m+(\kappa+k-\alpha m)e^{-2\kappa W}}
\,.
\end{equation}
Making a change of variable $k\to-k$ in the integral we find that the
surface charge density $\sigma_{-}$ is equal to the one computed in
Sec.~\ref{sec:inside-film} using the screening sum
rule~(\ref{eq:screen-sum-rule}), $\sigma_{-}=\sigma$, as it should
be. The screening sum rule~(\ref{eq:screen-sum-rule}) is then
verified.

\subsubsection{The correlations}

For both points $\r_1$ and~$\r_2$ on the boundary (region 1) we
compute the charge density correlation function by using
\begin{equation}
\label{eq:rho--deGreen-reg1}
\rho_{--}^{(2)T}(\r_1,\r_2)=-m_0^2 |g_{--}(\r_1,\r_2)|^2
.
\end{equation}
The Green function $g_{--}$ given by the Fourier transform of
Eq.~(\ref{eq:g--reg1}) is proportional to $m_0=\sqrt{mm_i}$. Then, the
charge density correlation function is proportional to $m_i^2$. This
gives a well defined surface charge density-surface charge density
correlation function
\begin{equation}
\left<\sigma_-(y_1)\sigma_-(y_2)\right>^T
=\delta^2 \rho_{--}^{(2)T}(\r_1,\r_2)
\,,
\end{equation}
in the limit $m_i\to\infty$, $\delta\to0$ and $m_i\delta =\alpha$
fixed. The final expression for the correlation function is
\begin{eqnarray}
\label{eq:sigma-sigma}
&&\left<\sigma_-(y_1)\sigma_-(y_2)\right>^T
=-
\left[\frac{m\alpha}{2\pi}\right]^2
\times\\
&&\times
\left|
\int_0^\infty 
\frac{(1-e^{-2\kappa W})e^{iky}\,dk}{%
\kappa+k+\alpha m+(\kappa-k-m\alpha)e^{-2\kappa W}}
\right|^2
,
\nonumber
\end{eqnarray}
with $y=y_1-y_2$.  As before it is interesting to study the decay of
the correlations along the $y$-axis. The discontinuity of the Fourier
transform of the Green function at $k=0$ will be translated into a
$1/y$ decay. Then the surface charge correlation will have the
asymptotic behavior
\begin{equation}
\label{eq:sigma-sigma-asymp}
\left<\sigma_{-}(y_1)\sigma_{-}(y_2)\right>^T
\simeq
-\frac{\alpha^2}{4\pi^2}
\frac{(1-e^{-2mW})^2}{\left(1+\alpha+(1-\alpha)e^{-2mW}\right)^2}
\frac{1}{y^2}
\,,
\end{equation}
and for very large films $W\to\infty$
\begin{equation}
\label{eq:asymp-sigmasigma}
\left<\sigma_-(y_1)\sigma_-(y_2)\right>^T
\simeq
-\frac{\alpha^2}{4\pi^2(\alpha+1)^2}
\frac{1}{y^2}
\,.
\end{equation}
In relation to the results of Sec.~\ref{sec:inside-film} on the
asymptotic behavior of the total charge correlation function
$S(\r_1,\r_2)$ inside the film we notice that for $y\to\infty$ and in
the limit $W\to\infty$ the following sum rule is verified
\begin{equation}
\left<\sigma_-(y_1)\sigma_-(y_2)\right>^T
=
\alpha^2
\int_0^\infty dx_2
\int_0^\infty dx_1
S(\r_1,\r_2)
\,.
\end{equation}

When $\r_1$ is in region 2 (inside the film) and $\r_2$ remains on the
boundary, the correlation function $\rho_{--}^{(2)T}$ is given by
\begin{equation}
\label{eq:rho--deGreen-reg2}
\rho_{--}^{(2)T}(\r_1,\r_2)=-m_0 m |g_{--}(\r_1,\r_2)|^2
,
\end{equation}
with the Green function $g_{--}$ given by
the inverse Fourier transform of Eq.~(\ref{eq:g--reg2}). The
correlation function $\rho_{+-}^{(2)T}$ is given by
\begin{equation}
\label{eq:rho+-deGreen-reg2}
\rho_{+-}^{(2)T}(\r_1,\r_2)=m_0 m |g_{+-}(\r_1,\r_2)|^2
,
\end{equation}
with $g_{+-}$ given by the inverse Fourier transform of
Eq.~(\ref{eq:g+-reg2}). 
We notice that both correlation functions are proportional to
$m_0^2=m_im$. Then the charge density-surface charge density
correlation $\left<\rho_{s}(\r_1)\sigma_{-}(y_2)\right>^T=\delta
\rho_{s-}^{(2)T}(\r_1,\r_2)$ will have a  well defined finite limit
when $m_i\to\infty$, $\delta\to0$ and $m_i\delta=\alpha$.

We find
\begin{eqnarray}
&&\left<\rho_{-}(\r_1)\sigma_{-}(y_2)\right>^T=
-
\frac{\alpha m^3}{\pi^2}\times
\\
&&
\times
\left|
\int_0^\infty
\frac{e^{-\kappa W} \sinh \left[\kappa(L-x_1)\right] e^{iky}\,dk}{%
\kappa+k+\alpha m+(\kappa-k-\alpha m)e^{-2\kappa W}}
\right|^2
,
\nonumber
\end{eqnarray}
and
\begin{eqnarray}
&&
\left<\rho_{+}(\r_1)\sigma_{-}(y_2)\right>^T=
\frac{\alpha m}{4\pi^2}
\times\\
&&
\times\left|
\int_0^\infty
\frac{e^{-\kappa(L+x_1)}
\left[\kappa+k+(\kappa-k)e^{-2\kappa(L-x_1)}\right]}{%
\kappa+k+\alpha m+(\kappa-k-\alpha m)e^{-2\kappa W}}
dk
\right|^2
.
\nonumber
\end{eqnarray}
It is interesting to notice a relation between $\left<\rho_{-}(\r_1)
\sigma_{-}(y_2)\right>^T$ and the surface charge density correlation
when both points are at the border, which we might call
``continuity''. This relation is
\begin{equation}
\left<\rho_{-}(x_1=-L,y_1)\sigma_{-}(y_2)\right>^T=
\frac{m}{\alpha}
\left<\sigma_-(y_1)\sigma_-(y_2)\right>^T
.
\end{equation}
It is also important to study the decay of the correlations along the
boundary, when $|y|=|y_1-y_2|\to\infty$. Here again the discontinuity of
the Fourier transform of the Green functions at $k=0$ is responsible
for a $1/y^2$ decay of the correlations. The asymptotic behavior of
the correlations for $|y_1-y_2|\to\infty$ is
\begin{equation}
\label{eq:rho-sigma-asymp}
\left<\rho_{-}(\r_1)\sigma_{-}(y_2)\right>^T\simeq
-
\frac{\alpha m\left[\sinh
m(L-x_1)\right]^2}{\left[1+\alpha+(1-\alpha)e^{-2mW}
\right]^2}
\frac{
e^{-2mW}}{\pi^2 y^2}
,
\end{equation}
and
\begin{equation}
\label{eq:rho+sigma-asymp}
\left<\rho_{+}(\r_1)\sigma_{-}(y_2)\right>^T\simeq
\frac{\alpha m\left[\cosh
m(L-x_1)\right]^2}{\left[1+\alpha+(1-\alpha)e^{-2mW}
\right]^2}
\frac{
e^{-2mW}}{\pi^2 y^2}
.
\end{equation}
For very large films $W\to\infty$, taking now the origin at the
boundary $x_1\to x_1+L$, the total structure function
$\left<\rho\sigma\right>^T=
\left<\rho_{-}\sigma_-\right>^T-\left<\rho_+\sigma_-\right>^T$ has the
asymptotic behavior
\begin{equation}
\label{eq:asymp-rhosigma}
\left<\rho(\r_1)\sigma(y_2)\right>^T\simeq 
-\frac{\alpha m
e^{-2mx_1}}{2\pi^2 (\alpha+1)^2 y^2}
\,.
\end{equation}
We notice that this asymptotic correlation function obeys a sum rule
with the charge correlation function $S(\r_1,\r_2)$ for both points
inside the fluid 
\begin{equation}
\int_0^\infty
S(\r_1,\r_2) \ dx_2
=
\alpha
\left<\rho(\r_1)\sigma(y_2)\right>^T
\,,
\end{equation}
and also a sum rule with the surface charge density correlation when
both points are on the boundary
\begin{equation}
\int_0^\infty
\left<\rho(\r_1)\sigma(y_2)\right>^T
\ dx_1
=
\alpha \left<\sigma(y_1)\sigma(y_2)\right>^T
\,.
\end{equation}

The remaining case to complete this study is when the two points are on
opposite boundaries, for instance $\r_2$ in region 1 and $\r_1$ in
region 3. In this case we found that all Green functions are of
order $O(m_0^{-1})$. Then, the correlation functions given by 
\begin{equation}
\rho_{ss'}^{(2)T}(\r_1,\r_2)=m_0^2 g_{ss'}(\r_1,\r_2)
g_{s's}(\r_2,\r_1)
\end{equation}
will be finite of order $O(1)$. But since there exists a surface
charge density at the boundaries, the interesting quantity here is the
surface charge correlation function
$\left<\sigma_s\sigma_{s'}\right>^T=\delta^2 \rho_{ss'}^{(2)T}$ which
will vanish in the limit $\delta\to 0$. We have the interesting
result: the surface charge densities at opposite boundaries are
completely uncorrelated.

To conclude this section let us return to the sum
rule~(\ref{eq:SumRule-orig}) for the correlations functions along the
boundary for $W\to\infty$. If one takes into account all contributions
(both particles in the fluid, one particle in the boundary and one in
the fluid and both particles in the boundary) to the charge density
correlation function, this function should read
\begin{subequations}
\label{eq:total-S}
\begin{eqnarray}
S_{\text{total}}(\r_1,\r_2)&=&S(\r_1,\r_2)
\label{eq:parte-S}
\\
\label{eq:parte-rhosigma-1}
&
+&
\delta(x_1)
\left<\sigma(x_1)\rho(\r_2)\right>^T
\\
\label{eq:parte-rhosigma-2}
&
+&
\delta(x_2)
\left<\sigma(x_2)\rho(\r_1)\right>^T
\\
\label{eq:parte-sigmasigma}
&+&
\delta(x_1)\delta(x_2)
\left<\sigma(x_1)\sigma(x_2)\right>^T
\end{eqnarray}
\end{subequations}
where the structure function $S(\r_1,\r_2)$ in Eq.~(\ref{eq:parte-S})
contains only the contributions for particles in the fluid, the
correlations $\left<\rho\sigma\right>^T$ in
Eqs.~(\ref{eq:parte-rhosigma-1}) and~(\ref{eq:parte-rhosigma-2})
contain the contribution when one point is on the boundary and
the other in the film, and finally the correlation
$\left<\sigma\sigma\right>^T$ contains the contribution when both
particles are on the boundary. The origin is taken on the boundary.
Using the asymptotic expressions for the different correlations given
by Eqs.~(\ref{eq:asymp-S}), (\ref{eq:asymp-rhosigma})
and~(\ref{eq:asymp-sigmasigma}) one can check that the sum
rule~(\ref{eq:SumRule-orig}) is verified:
\begin{eqnarray}
\int_0^\infty dx_1 \int_0^\infty dx_2
S_{\text{total}}(\r_1,\r_2) &=&
-\frac{1}{4\pi^2 y^2}
\frac{1+2\alpha+\alpha^2}{(\alpha+1)^2}
\nonumber
\\
&=&
-\frac{1}{4\pi^2 y^2}
\ 
.
\end{eqnarray}

For finite $W$ there exists also a sum rule similar to
Eq.~(\ref{eq:SumRule-orig}) for \textit{conducting} Coulomb
systems~\cite{MartinSumRules, JancoForresterSmith}. This sum rule reads
\begin{equation}
\label{eq:sum-rule-W-fini}
\int_{-L}^{L} dx_2 \int_{-L}^{L} dx_1
S_{\text{total}}(\r_1,\r_2)
=-\frac{1}{\beta \pi^2 y^2}
,
\end{equation}
for $|y|\to\infty$. From Eqs.~(\ref{eq:G--sol})
and~(\ref{eq:G++sol}) one can compute the asymptotic behavior of the
correlations for both points in the fluid. We find
\begin{equation}
S(\r_1,\r_2)\simeq
-\frac{m^2}{2\pi^2 y^2}
\frac{\cosh\left[2m(x_1+x_2)\right]}{\left[
\cosh(mW)+\alpha\sinh(mW)\right]^2}
\end{equation}
Taking into account all contributions from
Eqs.~(\ref{eq:sigma-sigma-asymp}), (\ref{eq:rho-sigma-asymp})
and~(\ref{eq:rho+sigma-asymp}) to the total charge correlation
function one finds
\begin{multline}
\int_{-L}^{L}\int_{-L}^{L}
S_{\text{total}}(\r_1,\r_2) dx_1 dx_2
=-
\frac{1}{2\pi^2 y^2}\times\\
\times\frac{\sinh^2 mW
+2\alpha\cosh mW \sinh mW
+\alpha^2 \sinh mW}{\left[\cosh mW +\alpha\sinh mW\right]^2}
\\
=-
\frac{1}{2\pi^2 y^2}
\left[1-\frac{1}{\left[\cosh mW +\alpha\sinh mW\right]^2}\right]
.
\end{multline}
The sum rule (\ref{eq:sum-rule-W-fini}) is not verified. The
discrepancy however is exponentially small when $W\to\infty$. This
situation also occurs in the case $\alpha=0$ which was studied in
Ref.~\cite{JancoManif}. The two-component plasma is no longer in its
conducting phase at $\Gamma=2$ when confined in a slab. This is a
special property of the two-component plasma at it is not related to
the short-range attractive potential near the boundaries. For this
reason the sum rule (\ref{eq:sum-rule-W-fini}) is no longer
valid. Actually for a Coulomb system in a dielectric phase, the
rhs.~of the sum rule~(\ref{eq:sum-rule-W-fini}) should read $(-1/\beta
\pi^2 y^2)(1-\epsilon^{-1})$ with $\epsilon$ the effective static
dielectric constant of the system. Then, in our case the system has a
effective dielectric constant given by
\begin{equation}
\epsilon=\left(\cosh mW +\alpha\sinh mW\right)^2
\end{equation}
This phenomenon is particular to the two-dimensional two-component
plasma and probably should not apply to real three-dimensional soap
films.

\section{Conclusion}
\label{sec:Conclusion}

We have studied a toy model for electrolytic soap films. Although this
model is very simple and gives only qualitative information for real
soap films it is very interesting since it is a solvable model. Our
study of the disjoining pressure shows that the charging of the
boundaries is responsible of the stability of the film. For strong
adhesivity $\alpha>1$ the film is stable while for weak adhesivity
$\alpha<1$ large films are not stable. For $0<\alpha<1$ large films
collapse to non-zero width small films. This could be the equivalent
of a collapse to a Common Black Films for our two-dimensional
model. For $\alpha=0$ unstable large films collapse to a film of zero
width which could be the equivalent of a Newton Black Film. We can
conclude that the Coulomb interaction plays indeed an important role
in the stability of large films. This is also the case in the
one-dimensional model presented in
Ref.~\cite{DeanHorganSentenac}. Then it is natural to expect that for
real three dimensional films the Coulomb interaction also plays an
important role in their stability. Of course in real films there are
other important interactions that certainly play a role in the
stability of the film and in particular in the stability and structure
of black films (both Common Black Films and Newton Black Films) that
have not been taken into account in our simplified model.

We also studied the density profiles and correlation functions for
this two-dimensional model. The density profile near a boundary shows
a classical double layered structure. A fraction of the anions (soap
molecules) are stucked in the boundary creating a first layer of
negative surface charge density. The ions in the fluid create the
second layer of positive charge and thickness given by the screening
length which screens the first layer.

The correlation functions exhibit the usual behavior. In the
$x$-direction across the film they decay exponentially with the
characteristic screening length. In the $y$-direction parallel to the
boundary they decay algebraically as $1/y^2$. The total charge
correlation function (taking into account all contributions of
particles in the fluid and in the boundary) obeys the usual sum rule
for Coulomb fluids near a plane wall when $W\to\infty$. For $W$
finite, the two-component plasma at $\Gamma=2$ is no longer a
conductor and therefore fails to satisfy a sum rule for correlations
along the boundaries. We also found an interesting new fact: the
surface charge densities on opposite boundaries are completely
uncorrelated.

\begin{acknowledgments}
G.~T.~would like to thank B.~Jancovici for useful discussions
concerning the sum rules of Sec.~\ref{sec:Density-corr}.  The
authors acknowledge partial financial support from COLCIENCIAS and BID
through project \# 1204-05-10078. G.~T.~acknowledge support from ECOS
Nord/ICFES action C00P02 of French and Colombian cooperation.
\end{acknowledgments}


\begin{thebibliography}{}
\bibitem{BelorgeyBenattar} O.~Belorgey and J.~J.~Benattar,
Phys.~Rev.~Lett.~\textbf{66}, 313 (1991).
\bibitem{SentenacBenattar} D.~Sentenac and J.~J.~Benattar,
Phys.~Rev.~Lett.~\textbf{81}, 160 (1998). 
\bibitem{DeanSentenac} D.~S.~Dean and D.~Sentenac,
Europhys.~Lett.~\textbf{38} 645 (1997)
\bibitem{DeanHorganSentenac} D.~S.~Dean, R.~R.~Horgan and D.~Sentenac,
J.~Stat.~Phys.~\textbf{90}, 899 (1998).
\bibitem{CornuJanco} F.~Cornu and B.~Jancovici,
J.~Chem.~Phys.~\textbf{90}, 2444 (1989).
\bibitem{JancoTellez-coulcrit} B.~Jancovici and G.~T\'ellez,
J.~Stat.~Phys.~\textbf{82}, 609 (1996).
\bibitem{Forrester} P.~J.~Forrester, J.~Stat.~Phys.~\textbf{67}, 433
(1992).
\bibitem{Tellez-tcp-neumann} G.~T\'ellez, J.~Stat.~Phys.~\textbf{104},
945 (2001).
\bibitem{JanManPis} B.~Jancovici, G.~Manificat and C.~Pisani,
J.~Stat.~Phys.~\textbf{76}, 307 (1994).
\bibitem{JancoSamaj-ideal-diel} B.~Jancovici and L.~\v{S}amaj,
J.~Stat.~Phys.~\textbf{104}, 755 (2001).
\bibitem{JancoManif} B.~Jancovici and G.~Manificat,
J.~Stat.~Phys.~\textbf{68}, 1089 (1992).
\bibitem{CornuThese} F.~Cornu, \textit{M\'ecanique Statistique Classique
de Syst\`emes Coulombiens Bidimensionels: Resultats Exacts},
Ph.~D.~Thesis, Orsay, France (1989).
\bibitem{JancoPlaneWall1} B.~Jancovici, J.~Stat.~Phys.~\textbf{28}, 43
(1982)
\bibitem{JancoPlaneWall2} B.~Jancovici, J.~Stat.~Phys.~\textbf{29}, 263
(1982)
\bibitem{MartinSumRules} Ph.~A.~Martin, Rev.~Mod.~Phys.~\textbf{60},
1075 (1988)
\bibitem{JancoForresterSmith} P.~J.~Forrester, B.~Jancovici and
E.~R.~Smith, J.~Stat.~Phys.~\textbf{31}, 129 (1983).
\end{thebibliography}
\end{document}